\documentclass[pra,aps,superscriptaddress,nofootinbib,twocolumn]{revtex4-1}
\usepackage{mathrsfs}
\usepackage{amsfonts}
\usepackage{amssymb}
\usepackage{amsmath}
\usepackage{graphicx}

\usepackage{appendix}

\usepackage[braket]{qcircuit}
\usepackage{algorithm}
\usepackage[noend]{algpseudocode}
\usepackage{algpseudocode}
\usepackage{subfigure,dcolumn}
\usepackage{graphicx}
\usepackage{mathtools}
\usepackage{braket}
\usepackage{subfigure}
\usepackage{epstopdf}
\usepackage[usenames,dvipsnames]{color}
\usepackage[colorlinks=true,citecolor=blue,linkcolor=magenta]{hyperref}
\usepackage{ulem}
\usepackage{lmodern}
\usepackage{amsthm}

\renewcommand{\i}{iH}

\newcommand{\Tr}{\mathrm{Tr}}

\newcommand{\0}{\text{ref}}

\newcommand{\ketbra}[2]{\vert #1 \rangle \langle #2 \vert}

\usepackage{listings}
\usepackage{algorithm}  
\usepackage{algpseudocode}  
\usepackage{amsmath}  

\usepackage[marginal]{footmisc}
\renewcommand{\thefootnote}

\begin{document}

\title{Quantum-enhanced Green's function Monte Carlo for excited states of nuclear shell model}

\author{Yongdan Yang}
\thanks{The two authors contributed equally to this work}
\affiliation{Graduate School of China Academy of Engineering Physics, Beijing 100193, China}

\author{Ruyu Yang}
\thanks{The two authors contributed equally to this work}
\affiliation{Graduate School of China Academy of Engineering Physics, Beijing 100193, China}

\author{Xiaosi Xu}
\email{xsxu@gscaep.ac.cn}
\affiliation{Graduate School of China Academy of Engineering Physics, Beijing 100193, China}

\begin{abstract}
We present a hybrid quantum-classical Green’s function Monte Carlo (GFMC) algorithm for estimating the excited states of the nuclear shell model. The conventional GFMC method, widely used to find the ground state of a quantum many-body system, is plagued by the sign problem, which leads to an exponentially increasing variance with the growth of system size and evolution time. This issue is typically mitigated by applying classical constraints but at the cost of introducing bias. Our approach uses quantum subspace diagonalization (QSD) on a quantum computer to prepare a quantum trial state, replacing the classical trial state in the GFMC process. We also incorporated a modified classical shadow technique in the implementation of QSD to optimize quantum resource utilization. Besides, we extend our hybrid GFMC algorithm to find the excited states of a given quantum system. Numerical results suggest our method largely enhances accuracy in determining excited state energies, offering an improvement over the conventional method.
\end{abstract}


\maketitle

\section{Introduction}
Finding the eigenvalues of quantum many-body systems is essential in various fields including quantum chemistry, drug discovery, nuclear physics, and condensed matter physics. However, direct diagonalization of a given Hamiltonian on a classical computer becomes increasingly difficult due to the exponentially increasing Hilbert space with the system's size. Various classical methods have been developed to handle this issue. However, popular methods such as the Density Matrix Renormalization Group (DMRG)~\cite{white1992density,white2004real,wouters2014density,baiardi2020density} and mean-field theory~\cite{kotliar2006electronic,negele1982mean} are either restricted to one-dimension systems or yield only approximate solutions. The Quantum Monte Carlo (QMC) methods can be applied to calculate large systems, while they suffer from the sign problem which has been proven to be NP hard~\cite{troyer2005computational}. Hence, developing a novel method to solve quantum many-body systems through ab initio calculations is a natural motivation. The Green’s function Monte Carlo (GFMC) method~\cite{van1994fixed,cheon1996green,sorella2000green,carlson2015quantum} belongs to the category of the QMC methods. By sampling a significant portion of the configuration space, GFMC can address certain quantum many-body problems with polynomial complexity. However, when applied to fermionic systems, GFMC faces the sign problem~\cite{troyer2005computational}, rendering the algorithm inefficient. Conventionally, we can handle this problem by introducing some constraints, for example as is done in the fixed-node GFMC (fnGFMC) method ~\cite{blunt2021fixed,ten1995proof}. Doing so leads to a bias in the final result. However, this bias can be reduced when the trial state approaches the exact eigenstate~\cite{xu2023quantum}. Besides, a recent work \cite{huggins2022unbiasing} demonstrated experimentally that a quantum trial state, generated by a quantum computer, can mitigate the bias when implementing auxiliary-field QMC to compute the ground state energy of many-electron systems. Therefore, the choice of trial state is significant to get a result with high accuracy.

The development of quantum computing provides a fundamentally different paradigm for performing calculations and offers us an alternative way to solve the quantum many-body Schrödinger equation from the first principle. Some quantum algorithms, e.g., quantum phase estimation (QPE)~\cite{nielsen2001quantum,kitaev1995quantum}, can compute the eigenstates of a quantum system by using quantum resources that scale polynomially with the system size.  However, this algorithm requires fault-tolerant quantum computers, which are still years away. Recently, several quantum algorithms based on the variational principle are developed to calculate the ground state of quantum many-body systems, e.g., the variational quantum eigensolver (VQE)~\cite{peruzzo2014variational,tilly2022variational,tang2021qubit,wang2019accelerated,kandala2017hardware}. These kind of algorithms are suitable for noisy intermediate-scale quantum (NISQ)~\cite{preskill2018quantum} devices and have proved successful in applications of a wide area. However, VQE suffers from the “barren-plateau”~\cite{mcclean2018barren,wang2021noise} problem when optimizing the parameters in classical computers. Besides, the effectiveness of VQE largely depends on the ansatz circuit, and a general way to construct an ansatz of high expressibility is still lacking. So, it's urged to develop non-variational algorithms with potential quantum advantages.

While substantial research has been directed toward solving the ground state problem of molecules~\cite{mcardle2020quantum,cao2019quantum}, there has been comparatively less focus on nuclear structure problems~\cite{jeukenne1976many,dumitrescu2018cloud} and determining the excited states. 
Calculating the ground and excited states of a nuclear system can reveal the properties of the nuclear force~\cite{roca2018nuclear},  the angular momentum, the parity, and the isospin of the nucleus. These properties determine how a nucleus interacts with other particles and how it can decay into another nucleus or emit radiation~\cite{wilson2021angular,ramsey2000nuclear,adelberger1985parity,van1996neutron,shin2016nuclear,sagawa1996effect,ejiri2000nuclear}. The study of nuclear-excited states can lay the foundation for future research on nuclear reaction processes.

In this paper, we propose a quantum-enhanced fnGFMC, which is a quantum-classical hybrid approach, to determine the excited state energies of the nuclear shell model (SM). To the best of our knowledge, no previous work has proposed a non-variational hybrid approach to address excited state problems in nuclear structures that surpasses the accuracy of classical methods. In our method, we apply the quantum subspace diagonalization (QSD) method~\cite{colless2018computation,yoshioka2022generalized,cortes2022quantum,stair2020multireference,francis2022subspace} on a quantum computer to find the trial state. To optimize quantum resource utilization, we employ the classical shadow technique~\cite{huang2020predicting} during QSD implementation, thereby creating a shadow version of the trial state. That means we only use the quantum computer in the stage of producing the trial state and the fnGFMC method is still implemented on a classical computer. Consequently, the evaluation of the overlap between the trial state and the random walker is precise.

To demonstrate the efficacy of our proposed algorithm, we performed calculations on the first excited state of certain nuclear systems, specifically $^{18}$O and $^{18}$F. The numerical results suggest that our method can significantly mitigate the bias introduced by the classical constraint in the fnGFMC approach. Additionally, we study the effect of Trotter error~\cite{trotter1959product,lloyd1996universal,berry2007efficient,wiebe2010higher} and the number of shots when constructing the classical shadow on the accuracy of our method. This analysis was carried out via numerical simulations, providing insights into the algorithm's performance under varying conditions.

This paper is structured as follows: Sec. \uppercase\expandafter{\romannumeral2} introduces the nuclear shell model. Sec. \uppercase\expandafter{\romannumeral3} provides an overview of the fnGFMC method. In Sec. \uppercase\expandafter{\romannumeral4}, we introduce the QSD method with classical shadow. The results of numerical simulations are shown in Sec.~\uppercase\expandafter{\romannumeral5}. Sec.~\uppercase\expandafter{\romannumeral6} analyzes the algorithmic error and the impact of shot noise. Sec.~\uppercase\expandafter{\romannumeral7} is the conclusion of this paper.

\section{The nuclear shell model}
In this section, we introduce the method to construct the nuclear shell model's Hamiltonians. Although the fundamental theory of QCD can interpret the origin of nuclear forces, these forces are often too complicated to precisely calculate. Thus, we can use a restricted Hilbert space and construct a phenomenological interaction that fits the experimental data. One example is the nuclear SM. The SM considers an inert core of closed shells, with the remaining valence nucleons filling the orbitals in the open shells. A single-particle state in SM has four good quantum numbers: the angular momentum $J$ and its 3rd component $M$, the isospin $T$, and its 3rd component $M_T$.

 In this paper, we demonstrate our method using the Wildental SM, also known as the USD SM~\cite{brown2006new,preedom1972shell,fiase1988effective}, which assumes a magic number nucleus that has a fixed core of ${}^{16}\mathrm{O}$ and valence nucleons in the sd-shell. The s-shell can accommodate 2 neutrons and 2 protons. The d-shell can accommodate 10 neutrons and 10 protons. Thus, this model applies to nuclei from ${}^{16}\mathrm{O}$ to ${}^{40}\mathrm{Ca}$.

The SM Hamiltonian consists of kinetic energy and two-body interactions. We label spherical orbitals with lowercase letters $a$, $b$, $c$, and $d$. We can write the Hamiltonian as
\begin{equation}
H^{f}=\sum_a \epsilon_a \hat{n}_a+\sum_{a \leqslant b, c \leqslant d} \sum_{J, T} V_{J T}(a b ; c d) \hat{T}_{J T}(a b ; c d),
\label{Eq:shell_model}
\end{equation}
where $\hat{n}_a$ denotes the occupation number for orbital $a$ with quantum numbers 
$n_a,l_a,m_a$ and $V_{J T}(a b ; c d)$ are parameters in this model. The Hamiltonian parameters $\epsilon_a$ and $V_{J T}(a b ; c d) $ are fitted to the experimental data. The scalar two-body density operator $\hat{T}_{J T}(a b ; c d)$ is given by
\begin{equation}
\hat{T}_{J T}(a b ; c d)=\sum_{M, M_T} A_{J M T M_T}^{\dagger}(a b) A_{J M T M_T}(c d).
\end{equation}
 The corresponding creation operator $A_{J M T M_T}^{\dagger}(a b)$ is given by
\begin{equation}
\begin{aligned}
    &A_{J M T M_T}^{\dagger}(a b) \equiv\left[c_a^{\dagger} \times c_b^{\dagger}\right]_{J M T M_T} = \\
&\sum_{m_a, m_b}\left(j_a m_a, j_b m_b \mid J M\right) \sum_{\mu_a, \mu_b}\left(\frac{1}{2} \mu_a, \frac{1}{2} \mu_b \mid T M_T\right)\\ &\hat{c}_{j_a m_a, \frac{1}{2} \mu_a}^{\dagger}
\hat{c}_{j_b m_b, \frac{1}{2} \mu_b}^{\dagger},
\end{aligned}
\end{equation}
where 
$(JM)$ and $(TM_T)$ denote the coupled spin and isospin quantum numbers, respectively. $j_a, j_b$ represent the angular momentum of uncoupled basis. $m_a, m_b$ refer to the 3rd component of angular momentum of uncoupled basis. $\mu_a, \mu_b$ are the 3rd component of isospin of uncoupled basis. $(\cdot|\cdot)$ denotes the Clebsch–Gordan coefficients.

\section{Green's function Monte Carlo method}
GFMC is a projector Monte Carlo technique that stochastically samples the ground-state wave function of a given Hamiltonian using the power method. The exact ground state $\ket{\varphi_g}$ can be extracted from the initial state $\ket{\varphi}$ that satisfies $\braket{\varphi|\varphi_g} \neq 0$ by~\cite{becca2017quantum}
\begin{equation}
    \lim_{k \to \infty}(\Lambda-H)^k \ket{\varphi} \propto \ket{\varphi_g}. 
\end{equation}
Here, $H$ denotes the Hamiltonian. $\Lambda$ is a large enough number that makes the diagonal elements of $G=\Lambda-H$ non-negative. For a sufficiently large $k$, we can derive the ground state energy $E_0$ as
\begin{equation}
\begin{aligned}
&E_0=\frac{\bra{\varphi_T}HG^k\ket{\varphi}}{\bra{\varphi_T}G^k\ket{\varphi}}\\
&=\frac{\sum\limits_{x_0,...,x_k}\bra{\varphi_T}H\ket{x_k}\bra{x_k}G\ket{x_{k-1}}...\bra{x_1}G\ket{x_0}\braket{x_0|\varphi}}{\sum\limits_{x_0,...,x_k}\braket{\varphi_T|x_k}\bra{x_k}G\ket{x_{k-1}}...\bra{x_1}G\ket{x_0}\braket{x_0|\varphi}},
\end{aligned}
\end{equation}
where ${\ket{x}}$ is a complete basis set in the Hilbert space. $\ket{\varphi_T}$ denotes the trial state, which can be different from the initial state $\ket{\varphi}$, and $G_{x x^{'}}=\bra{x}G\ket{x^{'}}$ is the so-called Green’s function. To estimate the first excited state, we can choose a trial state that has a finite overlap with the first excited state. To prevent the final result from collapsing to the ground state, we can use a filter operator on an initial guess state
\begin{equation}
    \ket{\varphi_T}=(I-\ket{\Tilde{\varphi}_g}\bra{\Tilde{\varphi}_g})\ket{\varphi_e}.
\end{equation}
Here, $\ket{\Tilde{\varphi}_g}\bra{\Tilde{\varphi}_g}$ is an approximation of the ground state. $\ket{\varphi_e}$ is an initial guess of the excited state. To obtain higher-energy excited states, one can follow a stepwise procedure by determining the approximation of the lower-energy energy states first and applying the filter operator: First, carefully select an excitation operator $U_2$ that makes the state $U_2\ket{HF}$ have a finite overlap with the second excited state. Then, use the operator $(I-\rho_E)(I-\rho_G)U_2\ket{HF}=\ket{\varphi}$ to filter out the components of the ground state and the first excited state, where $\rho_G$ and $\rho_E$ are the approximations of the ground state and first excited state, respectively, obtained from previous calculations. By repeating this process, we can evaluate other excited states as well. The sign of $G_{x x^{'}}$ is not always positive, which can cause the sign problem. We can constrain the sign problem with a classical technique, fixed-node GFMC (fnGFMC). In the fixed-node approximation, the Green’s function is~\cite{sorella1998green}
\begin{equation}
G_{x x^{'}}^{fn}=\left \{
\begin{aligned}
&-H_{x x^{'}}, \quad H_{x x^{'}}\leq 0 \text{ and } x\ne x^{'}\\
&\gamma H_{x x^{'}}, \quad H_{x x^{'}} > 0 \text{ and } x\ne x^{'}\\
&\Lambda-H_{x^{'} x^{'}}-(1+\gamma)V_{sf}(x^{'}),   \quad x=x^{'}.
\end{aligned}
\right.
\end{equation}
The diagonal sign-flip contribution $V_{sf}(x^{'})$ is given by
\begin{equation}
    V_{sf}(x^{'})=\sum_{x \neq x^{'},H_{x x^{'}}>0}H_{x x^{'}}.
\end{equation}
Here, $\gamma$ is a non-negative number between 0 and 1, and we set $\gamma=0$ in this paper. Although fix-node approximation can address the sign problem, it leads to a bias in the final result. In this work, we show in Sec.~\ref{sec:algorithm and numerical} that our quantum-enhanced fnGFMC method can effectively reduce the bias.

\section{Quantum Subspace Diagonalization method with classical shadow}

In this section, we introduce how to construct the quantum trial state $\ket{\varphi_T}$ for estimating the excited state in the GFMC method. For conventional GFMC, the trial state $\ket{\varphi_T}$ is usually restricted to states like Hartree-Fock state or a linear combination of mean-field states. However, those trial states may deviate significantly from the exact eigenstates, making the final estimation of eigenenergy inaccurate. The QSD method is a type of quantum algorithm that can solve eigenvalue problems on quantum computers. In this work, we apply QSD to obtain an approximation of the ground state, and then generate a trial state that has a finite overlap with the excited state (first excited state as an example shown in this paper). In QSD, we need to solve the generalized eigenvalue equation
\begin{equation} \label{geq}
H^s c=ESc,
\end{equation}
where $H_{ij}^s=\bra{\psi_i}H\ket{\psi_j}$ and $S_{ij}=\braket{\psi_i|\psi_j}$. $c$ denotes the generalized eigenstates. $\{\ket{\psi_i}\}$, $i=1,2,...,n$ form a basis of a subspace with dimension $n$, which can generally be much smaller than $N$, the dimension of the original Hamiltonian $H$. Hence, the generalized eigenvalue equation Eq.~\ref{geq} is solvable. There are three popular ways to generate the basis state $\ket{\psi_i}$, i.e., real-time evolution~\cite{cortes2022quantum, stair2020multireference}, imaginary-time evolution~\cite{kirby2023exact}, and quantum power method~\cite{seki2021quantum}. In our method, we utilize the quantum computer to generate the basis ${\ket{\psi_i}}$ by real-time evolution 
\begin{equation} \label{te}
 \ket{\psi_i}=e^{-iHt_i}\ket{\varphi},   
\end{equation}
where $t_i=i\Delta t$ with a fixed time step $\Delta t$ and $\ket{\varphi}$ represent an initial state that has a finite overlap with the ground state or excited state. Our algorithm employs the QSD method on a quantum computer to construct the trial state. In this paper, the evolution operator $e^{-iHt_i}$ is realized with Trotterization, a widely adopted approach. It is worth noting that there exist other algorithms, e.g., variational quantum simulation \cite{yuan2019theory}, which are more experimentally feasible alternatives for near-term quantum devices or early fault-tolerant quantum computers. 

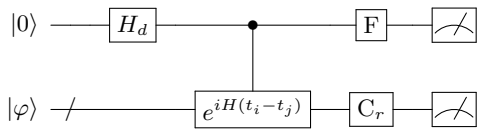
\begin{figure}
    \centerline{
\Qcircuit @C = 0.8em @R = 2em{
\lstick{\ket{0}} &\qw&\qw  &\gate{\text{$H_d$}} &\qw &\ctrl{1}&\qw &\gate{\text{F}}&\qw&\meter \\
\lstick{\ket{\varphi}} &{/}\qw&\qw &\qw&\qw &\gate{e^{iH(t_i - t_j)}} &\qw &\gate{\text{C}_r}&\qw&\meter\\
}
}
    \caption{Circuit to construct the classical shadow of $\rho_{ij}=\ket{\psi_i}\bra{\psi_j}$ where $i\neq j$. $H_d$ denotes the Hadamard gate. $F$ is either $H_d$ or $H_dF_S$ where $F_S$ is the phase gate diag(1,i). $C_r$ is a random local or global Clifford gate.}
    \label{fig: qubit efficient}
\end{figure}

When solving Eq.~\ref{geq}, some singular values of the matrix $S$ can be very small. As a result, the statistical fluctuations of the elements in matrices $H^s$ and $S$ can largely affect the results. Hence, we make use of the classical shadow technique introduced in Ref.~\cite{huang2020predicting}. However, if we directly use the classical shadow $\rho_i$ for each final basis state $\ket{\psi_i}$ and then to determine the matrix elements $H_{ij}^s$ and $S_{ij}$, the variance of results will increase exponentially with the system's size. This is because a non-linear function of shadow must be computed. Here, we introduce a novel way to construct the classical shadow to solve this problem. We assume the trial state we get from Eq.~\ref{geq} is
\begin{align}
    \ket{\varphi_T} = \sum_{i=1}^n c_i\ket{\psi_i},
\end{align}
where $\ket{\psi_i}$ is the basis state of the subspace and $c_i$ is the solution of Eq.~\ref{geq}. As mentioned before, when determining the matrix elements $H_{ij}^s$ and $S_{ij}$, we cannot use the classical shadow $\rho_i$ for each final basis state $\ket{\psi_i}$ directly. So, we use a simple quantum circuit in Fig.~\ref{fig: qubit efficient} to construct the classical shadow of $\rho_{ij}=\ket{\psi_i}\bra{\psi_j}$ where $i\neq j$. Note that the diagonal element of $H^s$ satisfy $\bra{\psi_i}H\ket{\psi_i}=\bra{\varphi}H\ket{\varphi}$ because the Hamiltonian is a conserved quantity (The details can be found in Appendix~\ref{sec A}). Then, the matrix elements $H_{ij}^s$ and $S_{ij}$ can be obtained by
\begin{align}\label{H_ij}
    H_{ij}^s=\Tr[H\rho_{ji}],\quad S_{ij}=\Tr[\rho_{ji}].
\end{align}
Here we apply the Bravyi-Kitaev transformation to convert the Hamiltonian $H$ into a form of Pauli operators. Since the Hamiltonian considered in our work is local, the number of terms in $H$ increases polynomially with the system size. Additionally, we can efficiently reconstruct the classical shadow of $\rho_{ji}$. Therefore, we can compute $\Tr[H\rho_{ji}]$ and $\Tr[\rho_{ji}]$ efficiently on a classical computer. In Fig.~\ref{fig: qubit efficient}, we implement $F=H_d$ and $F=H_d F_S$, where $H_d$ and $F_S$ are Hadamard gate and phase gate respectively, where $F_S$ is defined as 
\begin{align}
F_S = 
\begin{bmatrix}
1&0\\
0&i    
\end{bmatrix},
\end{align}
in the circuit in order to get the Hermitian and anti-Hermitian parts of the one-shot shadow. We discuss the variance of this method in Sec.\uppercase\expandafter{\romannumeral6}. 

\begin{figure*}[t]
\centering\includegraphics[width=1\hsize]{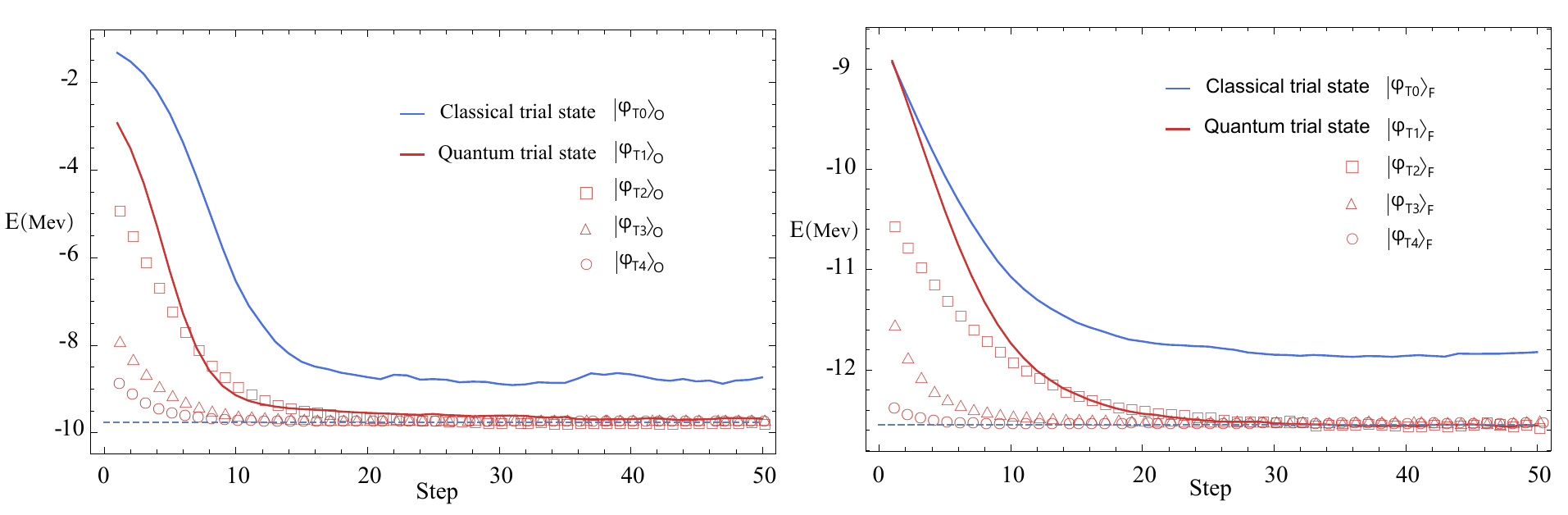}
\caption{The first excited state energy of the nuclear shell model $^{18}$O and $^{18}$F. The dashed blue line in the figures indicates the exact value of the first excited state energy, while the red curve, red square, red triangle, and red circle represent the results obtained by quantum-enhanced fnGFMC with different trial states. The blue curve represents the results obtained by conventional fnGFMC.}
\label{fig:1}
\end{figure*}

Then, we can solve the generalized eigenvalue Eq.~\ref{geq} to obtain $c_i$ on a classical computer and use the classical shadow $\rho_i$ of the basis state $\ket{\psi_i}$ to construct the trial state ${\rho_T}$ of the first excited state. The details are as follows:

\begin{enumerate}
    \item Select a state that has a reasonable overlap with the ground state, e.g., the Hartree Fock state $\ket{HF}$, which can be easily prepared on a quantum computer, as the initial state in Eq.~\ref{te}.
    \item Use the quantum circuit shown in Fig.~\ref{fig: qubit efficient} to construct the classical shadow $\hat{\rho}_{ij}$.
    \item Construct the matrices with $H_{ij}^s=\bra{\psi_i}H\ket{\psi_j}$ and $S_{ij}=\braket{\psi_i|\psi_j}$ using the classical shadow $\hat{\rho}_{ij}$ on a classical computer.
    \item Solve the generalized eigenvalue equation $H^sc=ESc$ on a classical computer and get the classical shadow ${\rho_G}$ as the approximation of the ground state.
    \item Select an initial state that has a reasonable overlap with the first excited state, e.g., $\ket{\varphi}=U_1\ket{HF}$. The operator $U_1$ represents an excitation operator that can make the state $U_1\ket{HF}$ have a reasonable overlap with the first excited state.
    \item Repeat the steps from 2 to 4 one more time with the initial state $U_1\ket{HF}$ prepared in step 5 to obtain the quantum trial state ${\rho_T}$ of the first excited state. Here, in step 3, we use the operator $(I-\rho_G)\ket{\psi_i}$ to filter out the component of the ground state on a classical computer, which makes ${\rho_T}$ orthogonal to the ground state. 
\end{enumerate}
In step 2, when we construct the classical shadow of the trial states, for a target error $\epsilon$, the required number of shots is bounded by $O(\frac{\| H\|^2_{shadow}}{\epsilon^2})$. Here, the shadow norm $\| H\|^2_{shadow}$ is defined as in Ref. \cite{huang2020predicting} and has been shown to increase polynomially with the system’s size, when $H$ is a local Hamiltonian as we consider in our work. Furthermore, the Trotter formula can be applied to decompose the controlled unitary operator $e^{iH(t_i-t_j)}$ implemented within the quantum circuit. As a result, simulating the circuit in Fig.~\ref{fig: qubit efficient} requires the number of quantum gates scaling polynomially with the system size and the evolution time $t_i-t_j$, with the latter upper bounded by the dimension $n$ of the subspace. Typically, this dimension is significantly smaller than the Hilbert space of the entire system, thereby not necessitating prolonged evolution time. As noted in Ref.~\cite{epperly2022theory}, the final error of the energy estimation using the QSD method is bounded by $O(p_0^{-2}\gamma^{-2n})$, where $p_0$ represents the overlap between the initial state and the target eigenstate, and $\gamma$ is a system-specific constant. This indicates that while the circuit depth increases proportionally with the subspace dimension, the accuracy of the results improves exponentially. In steps 3 and 4, the computational complexity is $O(n^2)$ and $O(n^3)$, where $n$ denotes the dimension of the subspace.
In step 5, we can set the excitation operator as $U_1=e^{a_i a_j ^ \dag - a_j a_i ^ \dag}$, where  $a_i$ and $a_i ^ \dag$ are the annihilation operator and the creation operator acting on the $i$-th qubit, respectively. Note that the anti-Hermitian operator $a_i a_j^\dag - a_j a_i^\dag$ is sparse, thus the excitation operator can be implemented efficiently using the methods outlined in Ref.~\cite{low2017optimal, berry2007efficient, berry2014exponential}. 

\section{The full algorithm and numerical simulations}\label{sec:algorithm and numerical}

In this section, we present the full algorithm of the quantum-enhanced fnGFMC approach and the results of numerical simulations. We take the first excited state as an example, while our method can be applied to find higher-level excited states. 


First, we combine the QSD method with classical shadow to create a shadow version of the trial state for evaluating the excited state. Using the trial
state ${\rho_T}$, we can evaluate the first excited state energy by
\begin{equation}\label{exf}
\begin{aligned}
&E_1= \\
&\frac{\sum\limits_{x_0,...,x_k}\Tr[\rho_T H \ket{x_k} \bra{\varphi_{\text{ref}}}] \bra{x_k}G\ket{x_{k-1}}...\bra{x_1}G\ket{x_0}\braket{x_0|\varphi}}{\sum\limits_{x_0,...,x_k}\Tr[\rho_T \ket{x_k} \bra{\varphi_\0}]\bra{x_k}G\ket{x_{k-1}}...\bra{x_1}G\ket{x_0}\braket{x_0|\varphi}}.
\end{aligned}
\end{equation}
Here, we introduce an auxiliary classical state $\ket{\varphi_\0}$, which can be an arbitrary state that has a finite overlap with $\ket{\varphi_T}$. By introducing this state we can obtain the inner product $\bra{\varphi_T}H\ket{x_k}$ and $\braket{\varphi_T|x_k}$, as
\begin{equation}\label{phase}
\begin{aligned}
&\Tr[\rho_T H \ket{x_k} \bra{\varphi_\0}]=\bra{\varphi_T}H\ket{x_k}\braket{\varphi_\0|\varphi_T}, \\
&\Tr[\rho_T \ket{x_k} \bra{\varphi_
\0}]=\braket{\varphi_T|x_k}\braket{\varphi_\0|\varphi_T}.
\end{aligned}
\end{equation}
Note that Eq.~\ref{phase} introduces an extra phase $\braket{\varphi_\0|\varphi_T}$. However, this does not affect the final result, as Eq.~\ref{exf} is a fractional formula. 

Next, we show the efficacy of our method by calculating the first excited state energy of the nuclear shell model $^{18}\text{O}$ and $^{18}\text{F}$ with 6 and 8 qubits. First, we use the conventional fnGFMC algorithm with a classical trial state to evaluate the first excited state energy. As a comparison, we utilize QSD to create several trial states for quantum-enhanced fnGFMC, simply by varying the time step $\Delta t$ in Eq.~\ref{te} when constructing the subspace.  We set the dimension of the subspace mentioned in Eq.~\ref{geq} as 4. Generally, the trial state gets closer to the first excited state with the increase of the subspace's dimension. The number of Monte Carlo sampling times is $10^6$. The results are shown in Fig.~\ref{fig:1}, which indicates that: (1) The excited state energy obtained from the conventional fnGFMC algorithm has a large bias from the exact value; However, this bias can be significantly reduced when applying the quantum-enhanced algorithm. (2) The trial states obtained from QSD can have different state fidelities with respect to the exact first excited state. Nevertheless, they all lead to results with a similarly higher level of accuracy than that from the conventional fnGFMC. Our numerical results demonstrate that, when having a similar overlap with the exact ground state, quantum trial states facilitate more accurate GFMC results.

\section{error analysis}
To generate the trial state with the QSD method, which is applied on a quantum computer, we use real-time evolution to construct the subspace and subsequently construct the classical shadow of the trial state.
The trial state is then used in conventional fnGFMC on classical computers to estimate the excited state. The error of our algorithm mainly stems from three aspects: (1) The statistical error resulted from a finite number of shots when constructing the classical shadow; (2) The Trotter error when realizing real-time evolution on a quantum computer; (3) The statistical error from Monte Carlo sampling when implementing fnGFMC. For (3), it's well known that the sampling noise of fnGFMC is proportional to $\frac{1}{\sqrt{N_s}}$, where $N_s$ denotes the total number of samples. Hence, we focus on (1) and (2) in this section.

\begin{figure}
\centering\includegraphics[width=1\hsize]{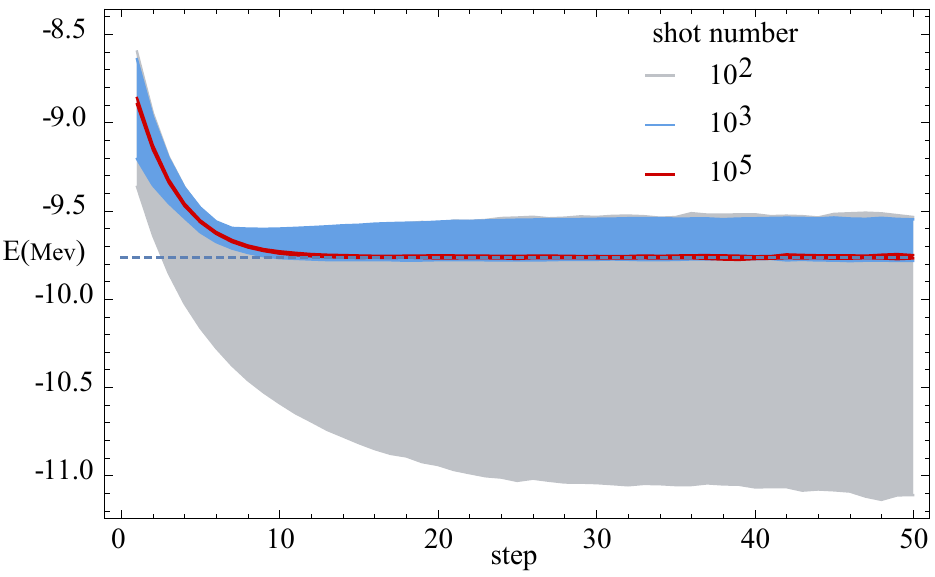}
\caption{The first excited state energy of the nuclear shell model $^{18}\text{O}$. We set different numbers of shots when constructing the classical shadow and fixed other parameters, e.g., evolution time and initial state. The dashed blue line in the figure shows the exact value of the first excited state energy. The gray region, blue region, and red region represent energies with different variances obtained with a shot number of $10^2$, $10^3$, and $10^5$ respectively.}
\label{fig:2}
\end{figure}

First, the shot number of classical shadow affects the accuracy of $H_{ij}^s$ and $S_{ij}$. Because the effect of shot number on $H_{ij}^s$ and $S_{ij}$ is the same, we only discuss the variance of the estimator $\hat{H}_{ij}^s = \Tr(\ket{\psi_i}\bra{\psi_j} H) $ from one-shot shadow $\ket{\psi_i}\bra{\psi_j}$, which is bounded by
\begin{align}
    \text{Var}(\hat{H}_{ij}^s) \leq 2\| H\|^2_{shadow}. 
\end{align}
\begin{figure}
\centering\includegraphics[width=1\hsize]{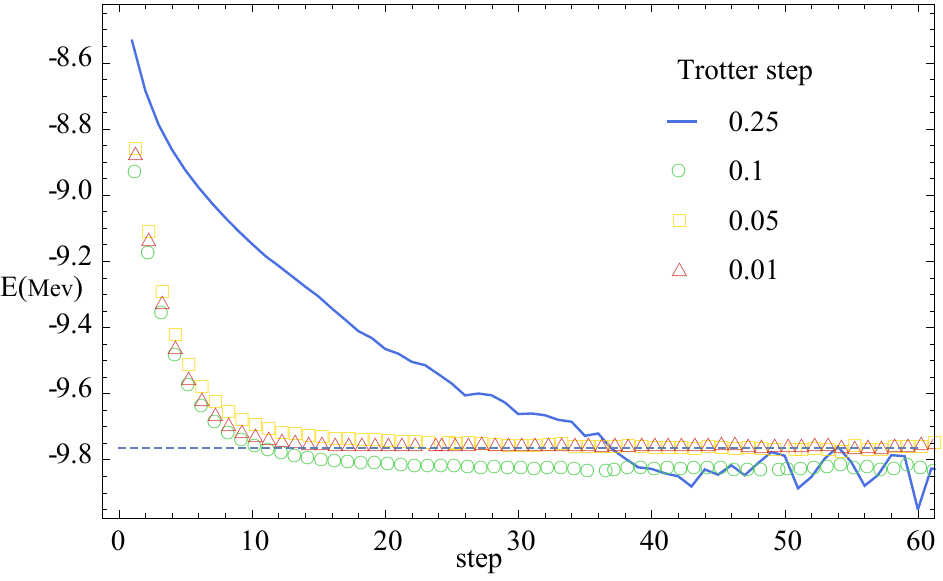}
\caption{The first excited state energy of the nuclear shell model $^{18}$O with different trotter steps when simulating the real-time evolution in QSD method. The dashed blue line in the figure indicates the exact value of the first excited state energy. The blue curve, green circle, yellow square, and red triangle are the results calculated by different Trotter steps.}
\label{fig:3}
\end{figure}
We define the shadow norm $\| H\|_{shadow}$ as in Ref~\cite{huang2020predicting}. The detailed calculation of the variance can be found in Appendix~\ref{sec: var}. The estimator from $N$-shot shadow has the variance 
\begin{align}
    \text{Var}(\hat{H}_{ij,N}^s) \leq \frac{2\| H\|^2_{shadow}}{N}. 
    \label{variance}
\end{align}
This indicates that the variance of the estimator $\hat{H}_{ij}^s$ decreases polynomially with the shot number N. Hence, we can efficiently calculate the matrix elements $H_{ij}^s$ and $S_{ij}$ to a desirable accuracy. The overall cost of our quantum-enhanced fnGFMC approach comprises the costs of classical shadow $O(\frac{\| H\|^2_{shadow}}{\epsilon^2})$ and Monte Carlo $O(\frac{1}{\epsilon_n^2})$, where $\epsilon$ and $\epsilon_n$ represent the desired accuracy of classical shadow and Monte Carlo method respectively. In Fig.~\ref{fig:2}, we vary the shot number from $10^2$ to $10^5$ when constructing the classical shadow. The results converge when the shot number is larger than $10^5$. 

In this work, we employ Trotterization to implement the real-time evolution operator $e^{-iHt}$ when using the QSD method. When given a Hamiltonian $\sum_{j=1}^{M}H_j$, where $H_j$ are non-commutive terms, the first order Trotter formula reads
\begin{equation} \label{tr}
    e^{-iH\Delta t}=e^{-i H_M \Delta t}...e^{-i H_1 \Delta t}+O(\Delta t^2 ).
\end{equation}
In the QSD method, we can expand the trial state $\ket{\varphi_T}$ as 
\begin{equation} \label{ex}
    \ket{\varphi_T}=\sum_{i=1}^n c_i (e^{-iH \Delta t})^{i}\ket{\varphi},
\end{equation}
where $\ket{\varphi}$ is the initial state and $i\Delta t=t_i$, $i=1,2,...,n$. The amplitude $c_i$ can be calculated by Eq.~\ref{geq}. Then, we can substitute Eq.~\ref{tr} into Eq.~\ref{ex} and derive the error caused by the expansion as
\begin{equation}
\begin{aligned}
     &\ket{\varphi_T}-\ket{\varphi_T}^{err} \\
    =&\sum_{i=1}^n c_i [(e^{-iH \Delta t})^{i}-(e^{-i H_M \Delta t}...e^{-i H_1 \Delta t})^{i}]\ket{\varphi} \\
    \approx&O(\Delta t^2 )\sum_{i=1}^n i c_i (e^{-i H_M \Delta t}...e^{-i H_1 \Delta t})^{i-1}\ket{\varphi}.
\end{aligned}    
\end{equation}
We see the Trotter expansion introduces an error of order $O(\Delta t^2 )$. In Fig.~\ref{fig:3}, we chose different Trotter steps ranging from 0.01 to 0.25 while simulating the real-time evolution for constructing the subspace basis. We set the shot number as $10^5$ here. As the Trotter step gets smaller, the effect of Trotter error diminishes as we get an energy closer to the exact value.

\section{conclusion}
In this paper, we introduce a novel quantum-enhanced fnGFMC method for estimating the excited states of nuclear shell models. This approach marks a significant advancement over the conventional fnGFMC method by leveraging QSD to generate a quantum trial state in place of the classical one. The primary advantage of this substitution is the reduction in bias induced by classical constraint techniques, for example, the fixed-node approximation in this paper. Furthermore, we incorporated the classical shadow technique in the implementation of QSD, optimizing the quantum measurement process during the evaluation of overlaps between the trial and basis states. Our method was tested through numerical simulations on various nuclear system models, including $^{18}$O and $^{18}$F. The results demonstrate the enhanced accuracy of our method in estimating excited states, as compared to its classical counterparts. This accuracy is crucial for in-depth studies in nuclear physics, offering a more reliable tool for understanding complex nuclear interactions and properties. 

In summary, our research not only demonstrates the potential of the quantum-enhanced fnGFMC method in accurately estimating excited states but also opens new avenues for research in nuclear physics, particularly in the study of excited states.

\begin{acknowledgments}
We thank fruitful discussions with Zongkang Zhang.
This work is supported by National Natural Science Foundation of China (Grant No. 12088101, 12225507, 12347124) and NSAF (Grant No. U1930403, U2330201 and U2330401).
\end{acknowledgments}

\bibliography{ref.bib}
\onecolumngrid
\appendix

 \section{Construction of the classical shadow}
 \label{sec A}
In this section, we give the details of constructing the classic shadow. 
The circuit is shown in Fig~\ref{fig: qubit efficient}. The state before the random Clifford gate $C_r$ and $F$ is 
\begin{align}
     & \frac{1}{2}\ketbra{0}{0} \otimes \ketbra{\varphi}{\varphi} + \frac{1}{2}\ketbra{0}{1}\otimes\ketbra{\varphi}{\varphi} e^{-iH(t_i-t_j)}\nonumber\\  +&\frac{1}{2} \ketbra{1}{0}\otimes e^{iH(t_i-t_j)}\ketbra{\varphi}{\varphi}  +\frac{1}{2} \ketbra{1}{1}\otimes e^{iH(t_i-t_j)}\ketbra{\varphi}{\varphi}  e^{-iH(t_i-t_j)}.
\end{align}
Here, note that $H_{i,j}^s=Tr[He^{iH(t_i-t_j)}\ketbra{\varphi}{\varphi}]$. Apply Hadamard gate $H_d$ on the ancilla qubit and we obtain
\begin{align}
    & \frac{1}{4}(\ketbra{0}{0} + \ketbra{0}{1} + \ketbra{1}{0} + \ketbra{1}{1}) \otimes \ketbra{\varphi}{\varphi}  + \frac{1}{4}(\ketbra{0}{0} - \ketbra{0}{1} + \ketbra{1}{0} - \ketbra{1}{1})\otimes\ketbra{\varphi}{\varphi} e^{-\i(t_i-t_j)}\nonumber\\  +& \frac{1}{4}(\ketbra{0}{0} + \ketbra{0}{1} - \ketbra{1}{0} - \ketbra{1}{1} )\otimes e^{\i(t_i-t_j)}\ketbra{\varphi}{\varphi}  + \frac{1}{4}(\ketbra{0}{0} - \ketbra{0}{1} - \ketbra{1}{0} + \ketbra{1}{1} )\otimes e^{\i(t_i-t_j)}\ketbra{\varphi}{\varphi} e^{-\i(t_i-t_j)}\nonumber\\
     = &\frac{1}{4}\ketbra{0}{0}\otimes(\ketbra{\varphi}{\varphi} + \ketbra{\varphi}{\varphi}e^{-\i(t_i-t_j)} + e^{\i(t_i-t_j)}\ketbra{\varphi}{\varphi} + e^{\i(t_i-t_j)}\ketbra{\varphi}{\varphi}e^{-\i(t_i-t_j)})\nonumber\\
    + &\frac{1}{4}\ketbra{0}{1}\otimes(\ketbra{\varphi}{\varphi} - \ketbra{\varphi}{\varphi}e^{-\i(t_i-t_j)} + e^{\i(t_i-t_j)}\ketbra{\varphi}{\varphi} - e^{\i(t_i-t_j)}\ketbra{\varphi}{\varphi}e^{-\i(t_i-t_j)})\nonumber\\
     + &\frac{1}{4}\ketbra{1}{0}\otimes (\ketbra{\varphi}{\varphi} + \ketbra{\varphi}{\varphi}e^{-\i(t_i-t_j)} - e^{\i(t_i-t_j)}\ketbra{\varphi}{\varphi} - e^{\i(t_i-t_j)}\ketbra{\varphi}{\varphi}e^{-\i(t_i-t_j)})\nonumber\\
     + &\frac{1}{4}\ketbra{1}{1}\otimes(\ketbra{\varphi}{\varphi} - \ketbra{\varphi}{\varphi}e^{-\i(t_i-t_j)} - e^{\i(t_i-t_j)}\ketbra{\varphi}{\varphi} + e^{\i(t_i-t_j)}\ketbra{\varphi}{\varphi}e^{-\i(t_i-t_j)}).
\end{align}
Apply measurement on the ancilla qubit. If we obtain $\ket{0}$, the target system is
\begin{align}
&\frac{1}{4}(\ketbra{\varphi}{\varphi} + \ketbra{\varphi}{\varphi}e^{-iH(t_i-t_j)} + e^{iH(t_i-t_j)}\ketbra{\varphi}{\varphi} + e^{iH(t_i-t_j)}\ketbra{\varphi}{\varphi}e^{-iH(t_i-t_j)})\\ 
&= \frac{1}{4}( \ket{\varphi} + e^{iH(t_i-t_j)}\ket{\varphi})(\bra{\varphi} + \bra{\varphi}e^{-iH(t_i-t_j)}).
\end{align}
Then we can construct the classical shadow $\hat{\rho}_{+}$ of the unnormalized state $\frac{1}{4}( \ket{\varphi} + e^{iH(t_i-t_j)}\ket{\varphi})(\bra{\varphi} + \bra{\varphi}e^{-iH(t_i-t_j)})$.
If we obtain $\ket{1}$, the target system is 
\begin{align}
&\frac{1}{4}(\ketbra{\varphi}{\varphi} - \ketbra{\varphi}{\varphi}e^{-iH(t_i-t_j)} - e^{iH(t_i-t_j)}\ketbra{\varphi}{\varphi} + e^{iH(t_i-t_j)}\ketbra{\varphi}{\varphi}e^{-iH(t_i-t_j)})\\ 
&= \frac{1}{4}( \ket{\varphi} - e^{iH(t_i-t_j)}\ket{\varphi})(\bra{\varphi} - \bra{\varphi}e^{-iH(t_i-t_j)}).
\end{align}
Then we can construct the classical shadow $\hat{\rho}_{-}$ of the unnormalized state $\frac{1}{4}( \ket{\varphi} - e^{iH(t_i-t_j)}\ket{\varphi})(\bra{\varphi} - \bra{\varphi}e^{-iH(t_i-t_j)})$. As a result, we can construct the shadow $\hat{\rho}^R = \hat{\rho}_{+} - \hat{\rho}_{-}$ of the operator $(\ketbra{\varphi}{\varphi}e^{-iH(t_i-t_j)} + e^{iH(t_i-t_j)}\ketbra{\varphi}{\varphi})/2$. One can verify that 
\begin{align}
    \mathbf{E}[\Tr(H\hat{\rho}^R)] =\frac{1}{2} \Tr(H(\ketbra{\varphi}{\varphi}e^{-\i(t_i-t_j)} + e^{\i(t_i- t_j)}\ketbra{\varphi}{\varphi}))=Re[H_{ij}^s].
\end{align}
Setting $F$ to $H_dG_S$, we can obtain the classical shadow $\hat{\rho}^I$ of the operator $(\ketbra{\varphi}{\varphi}e^{-\i(t_i-t_j)} - e^{\i(t_i-t_j)}\ketbra{\varphi}{\varphi})/2i$ and have
\begin{align}
    \mathbf{E}[\Tr(H\hat{\rho}^I)] =\frac{-i}{2} \Tr(H(\ketbra{\varphi}{\varphi}e^{-\i(t_i-t_j)} - e^{\i(t_i- t_j)}\ketbra{\varphi}{\varphi}))=Im[H_{ij}^s].
\end{align}

We summarize the protocol as follows:
\begin{enumerate}
    \item Set $F = H_d$($F = H_dF_s$), and $\hat{\rho}^R = 0$($\hat{\rho}^I$ =0).
    \item Randomly choose a global or local Clifford gate $C_r$.
    \item Apply $Z$ measurement on all qubits and obtain the result $\ket{z_0}\otimes \ket{z}$.
    \item If $z_0$ = 0:
    \subitem Update $\hat{\rho}^R$($\hat{\rho}^I$) as $\hat{\rho}^R + M^{-1}(C_r\ketbra{z}{z}C_r^{\dagger}$) ($\hat{\rho}^I + M^{-1}(C_r\ketbra{z}{z}C_r^{\dagger}$ )).

    If $z_0 = 1$:
        \subitem Update $\hat{\rho}^R$($\hat{\rho}^I$) as $\hat{\rho}^R - M^{-1}(C_r\ketbra{z}{z}C_r^{\dagger}$ )($\hat{\rho}^I - M^{-1}(C_r\ketbra{z}{z}C_r^{\dagger}$ )),

where $M^{-1}$ is the linear transformation depending on the ensemble of $C_r$. If $C_r$ is an $n$-qubit global Clifford gate, 
\begin{align}
M^{-1}(X) = (2^n+1)O - I_n,
\end{align}
where $I_n$ is a $2^n\times 2^n$ identity matrix, and $X$ is a density matrix. When $C_r$ is the tensor product of single-qubit Clifford gates,
\begin{align}
    M^{-1}(X) = \bigotimes_{i=1}^n M_1^{-1}(X),
\end{align}
where $M_1^{-1}(A) = 3A - I_1$ for arbitrary single qubit density operator $A$.
    \item Repeat steps 2-4 $N$ times and take the average $\hat{\rho}^R\leftarrow \hat{\rho}^R/N$ ($\hat{\rho}^I\leftarrow \hat{\rho}^I/N$).
\end{enumerate}
The full result is given by 
\begin{align}
    \mathbf{E}[\Tr(H(\hat{\rho}^R + i\hat{\rho}^I))] =H_{ij}^s.
\end{align}

\section{Variance of the shadow estimation}
\label{sec: var} 
In this section, we show how to derive the variance. The variance of one-shot shadow estimation can be bounded by 
\begin{align}
    \mathbf{Var}[\Tr(O\hat{\rho}^R) + i\Tr(O\hat{\rho}^I)] \leq \mathbf{Var}[\Tr(O\hat{\rho}^R)] + \mathbf{Var}[\Tr(O\hat{\rho}^I)].
\end{align}
In our algorithm, $O$ can be $H$ and $I$.
From the definition of variance, we obtain that
\begin{align}
    &\mathbf{Var}[\Tr(O\hat{\rho}^R)]\nonumber\\ 
\leq & \mathbf{E}[|\Tr(O\hat{\rho}^R)|^2]\nonumber\\
 = & \mathbf{E}[| \bra{z} C_r M^{-1}(O)C_r^{\dagger}\ket{z}  |^2 ]\nonumber\\
 = & p_{+}\mathbf{E}_{C_r}\left[\sum_{z} \bra{z}C_r\rho_{+}C_r^{\dagger}\ket{z}  \left| \bra{z} C_r M^{-1}(O)C_r^{\dagger}\ket{z}  \right|^2 \right] + p_{-}\mathbf{E}_{C_r}\left[\sum_{z} \bra{z}C_r\rho_{-}C_r^{\dagger}\ket{z}  \left| \bra{z} C_r M^{-1}(O)C_r^{\dagger}\ket{z}  \right|^2 \right]\nonumber\\
 = & \mathbf{E}_{C_r}\left[\sum_{z} \bra{z}C_r (p_{+}\rho_{+} +p_{-}\rho_{-})C_r^{\dagger}\ket{z}  \left| \bra{z} C_r M^{-1}(O)C_r^{\dagger}\ket{z}  \right|^2 \right] 
\end{align}
where $p_{+}$ and $p_{-}$ are the probability of obtaining $z_0 = 0$ and $z_0=1$ on the ancilla qubit, respectively. $\rho_{+}= \mathbf{E}[\hat{\rho}^R]$ and $\rho_{-}=\mathbf{E}[\hat{\rho}^I]$ are the reduced density matrix after measuring the ancilla qubit and before applying the randomized Clifford gate $C_r$. 
Since $p_{+} + p_{-} = 1$, the state $p_{+}\rho_{+} + p_{-}\rho_{-}$ is a valid quantum state.  Hence, the variance can be upper bounded by the shadow norm of the observable:
\begin{align}
\mathbf{Var}[\Tr(O\hat{\rho}^R)] & \leq \max_{\sigma: state} \mathbf{E}_{C_r}\left[\sum_{z} \bra{z}C_r\sigma C_r^{\dagger}\ket{z}  \left| \bra{z} C_r M^{-1}(O)C_r^{\dagger}\ket{z}  \right|^2 \right]\nonumber\\
&= \|O\|^2_{shadow},
\label{shadow norm}
\end{align}
where $M^{-1}$ is the linear transformation depending on the ensemble of $C_r$ and the last line of Eq.~\ref{shadow norm} comes from the definition of shadow norm. 
If $C_r$ is an $n$-qubit global Clifford gate, 
\begin{align}
M^{-1}(O) = (2^n+1)O - \Tr(O)I_n,
\end{align}
where $I_n$ is a $2^n\times 2^n$ identity matrix. When $C_r$ is the tensor product of single-qubit Clifford gates,
\begin{align}
    M^{-1}(O) = \bigotimes_{i=1}^n M_1^{-1}(O),
\end{align}
where $M_1^{-1}(A) = 3A - \Tr(A)I_1$ for arbitrary single qubit operator $A$.
Similarly, the same way can be used for the non-Hermitian part
\begin{align}
     \mathbf{Var}[\Tr(O\hat{\rho}^I)] \leq \|O\|^2_{shadow}.
\end{align}
Thus, the variance of the full estimation can be bounded as
\begin{align}
     \mathbf{Var}[\Tr(O\hat{\rho}^R) + i\Tr(O\hat{\rho}^I)] \leq 2\|O\|^2_{shadow}.
\end{align}

\end{document}